\documentclass[a4paper,10pt]{article}
\usepackage{amsmath,amsfonts}
\usepackage{times}
\usepackage[T1]{fontenc}
\usepackage{epsfig}
\usepackage{amssymb}
\usepackage[english]{babel}

\usepackage[usenames,dvipsnames]{color}
\usepackage{color}

\begin{document}

\title{\Huge{\bf A BPS Skyrme model and baryons at large $N_c$}}

\author{C. Adam$^{a)}$\thanks{adam@fpaxp1.usc.es},
J. S\'{a}nchez-Guill\'{e}n
$^{a)b)}$\thanks{joaquin@fpaxp1.usc.es}, A. Wereszczy\'{n}ski
$^{c)}$\thanks{wereszczynski@th.if.uj.edu.pl}
\\ \\ 
$^{a)}$ Departamento de F\'isica de Part\'iculas, Universidad
\\ 
de Santiago, and Instituto Galego de F\'isica de Altas Enerx\'ias
\\ 
(IGFAE) E-15782 Santiago de Compostela, Spain
\\ \\ 
$^{b)}$ Sabbatical leave at: Departamento de F\'isica Te\'orica,
\\ 
Universidad de Zaragoza, 50009 Zaragoza, Spain
\\ \\
$^{c)}$ Institute of Physics,  Jagiellonian University,
\\ 
Reymonta 4, Krak\'{o}w, Poland}

\maketitle

\begin{abstract}
Within the class of field theories with the field contents of the Skyrme
model, one submodel can be found which consists of the square of the baryon
current and a potential term only. For this submodel, a Bogomolny bound exists
and the static soliton solutions saturate this bound. Further, already on the
classical level, this BPS Skyrme model reproduces some features of the 
liquid drop model
of nuclei. Here, we investigate the model in more detail and, besides, we
perform the rigid rotor quantization of the simplest Skyrmion (the nucleon).
In addition, we discuss indications that the viability of the model
as a low energy effective field theory for QCD is further improved
in the limit of a large number of colors $N_c$.

\end{abstract}

\newpage
%%%%%%%%%%%%%%%%%%%%%%%%%%%%%%%%%%%%%%%%%%%%%%%%%%%%%%%%%%%%%%
\section{Introduction}
%%%%%%%%%%%%%%%%%%%%%%%%%%%%%%%%%%%%%%%%%%%%%%%%%%%%%%%%%%%%%
The derivation of the correct description of low energy hadron dynamics 
is undoubtedly one of the most prominent challenges of QCD. The difficulty 
of the problem originates in the non-perturbative nature of the quark
and gluon
interactions in the low energy limit. In the large $N_c$ limit, however, 
it is known that QCD becomes equivalent to an effective theory of mesons 
\cite{thooft}. Baryons (hadrons as well as atomic nuclei) appear as 
solitonic excitations with an identification between the baryon number 
and the topological charge \cite{witten1}.  One of the most popular 
realizations of this idea is the Skyrme model \cite{skyrme} i.e., a 
version of a phenomenological chiral Lagrangian, where the primary 
ingredients are meson fields. 
Static properties of baryons as well as nuclei are derived with the help of 
the semiclassical quantization of the solitonic zero modes \cite{nappi1}, 
\cite{nappi2}, \cite{sutcliffe1}. While many phenomenological properties 
of baryonic (and nuclear) matter seem to fit perfectly in the framework of 
the Skyrme theory, there are still some results which are in 
disagreement with the experimental or lattice data. 
\\
First of all, there is a conceptual problem with the
derivation of the Skyrme 
model from the underlying quantum theory. Assuming the large $N_c$ expansion, 
one may derive the pure pseudscalar low-energy dynamics from QCD in the 
leading order in derivatives usually truncated at 4th \cite{simic}, 
\cite{aitchison}, or 6th \cite{zuk}, \cite{fraser}, \cite{aitchison2}
order terms. However, baryons being solitons i.e., 
extended collective excitations of the chiral field, they contain regions with 
rather large values of the gradients of the field. Thus, any truncation does 
not seem to be justified by a small derivative expansion and one should 
instead consider more terms \cite{adkins}- \cite{vec skyrme} (also an 
infinite series of terms \cite{marleau}) or find another acceptable 
principle which could provide a selection criterium for 
a simple effective action. 
The situation is further complicated by the fact that, 
already at the 4th order in the derivative 
expansion, one gets not only the standard Skyrme term but also two additional 
terms which contribute at the same level to the effective action. In other 
words, there is no reason to omit them in this kind of expansion. Moreover, 
one of these terms contains the second time derivative squared and, therefore, 
leads to serious problems in the dynamics (the standard Cauchy data are not 
sufficient to determine the time evolution uniquely) as well as in the 
quantization procedure (no obvious Hamiltonian). Additionally, this term enters 
with a wrong sign, destabilizing soliton configurations. All this seems 
to indicate that a
simultaneous large $N_c$ and small derivative 
expansion might be problematic. 
\\
Another source of criticism of the Skyrme model is related to certain 
phenomenological features of solutions of the model \cite{larry-1},
\cite{larry}, \cite{torri}. 
\begin{itemize}
\item[] {\it Large binding energies.} As the Skyrme model is not an exact BPS 
theory, its soliton solutions do not saturate the corresponding 
linear energy-topological charge relation which results in the appearance of 
binding energies. Unfortunately, their value is significantly bigger than 
experimental energies which do not exceed 1\% of the nuclei masses 
\cite{massless}-\cite{sutcliffe bps}. From the point of view 
of the large $N_c$ 
expansion this seems to indicate that the binding energies scale like 
$N_c\Lambda_{\mbox{QCD}}$, instead of $\Lambda_{\mbox{QCD}}/N_c$ 
as expected for 
the weakly bound nuclear matter. 
\item[] {\it Crystal state of matter.} The matter described by the Skyrme 
model in the limit of large baryon charge
behaves like a crystal, not as a liquid for $N_c \rightarrow \infty$ 
\cite{klebanov} as well as for finite $N_c$ \cite{massless}-\cite{manton}. 
Moreover, shell-like structures 
are preferred rather than core or ball-type configurations. 
This can be improved 
by including the potential term (that is massive pions), but still for a 
fixed value of the 
mass parameter, the first few skyrmions possess shell-like structures.  
\item[] {\it Strong forces at intermediate distances.} Due to the enhancement 
of the pion coupling constant in the Skyrme model $g_{\pi NN} \sim N_c^{3/2}$, 
the axial coupling constant grows linearly with $N_c$. This leads to strong 
spin-isospin forces at distances larger than the size of nucleus, which is in 
contradiction to experimental as well as lattice data \cite{lattice}.
\end{itemize}
Let us remark that in 
Ref. \cite{larry} a critical evaluation of large $N_c$ properties has been 
conducted not only for the standard Skyrme model but also for the standard
non-relativistic quark model. In the latter case, the solution proposed in
\cite{larry} consists in the  so-called dichotomous 
nucleon model, where one assumes that for odd $N_c$  
 all quarks are paired in diquarks except for one which carries the quantum 
numbers 
of the nucleon and has much larger spatial extension than the diquarks. 
The resulting very small overlap in wavefunctions (``dichotomy'') tames the
strong forces at large $N_c$. 
\\
The problems mentioned above 
may suggest that the Skyrme model (at least in the usually 
considered versions) probably under certain circumstances 
is not the right starting point for the effective 
low-energy description of baryons for large $N_c$.  
\\
On the other hand, model independent results, which are related to 
topological 
and geometrical aspects of the solutions rather than to a particular form of 
the action, may indicate the correctness of the original 
topological concept of Skyrme. 
One can still expect that mesonic matrix fields (or possibly their
generalizations) are the
right low energy degrees of freedom (in agreement with \cite{thooft}), 
even when the proper effective action is not accessible via the small 
derivative expansion. This motivates the philosophy of our approach. We use 
the SU(2) matrix field, i.e., keep the topological contents of the Skyrme 
model but change the Lagrangian. In fact, the unique principles we are left 
with if we do not want to rely on the derivative expansion are, 
again, topology and the need for a 
BPS theory with chiral skyrme fields. 
\\
The BPS Skyrme model we propose \cite{proposal}
is by construction even more topological 
in nature than any of the standard versions of the Skyrme model. 
It shares with the standard Skyrme model the stabilization of soliton
solutions by a higher order term in derivatives (a sextic term in our case).
It is an 
example of a BPS theory with topological solitons saturating the pertinent 
Bogomolny bound and therefore, with zero binding energies. Further, this 
BPS model possesses a huge number of symmetries which lead to its 
integrability. What is more important, among its symmetries are the volume 
preserving diffeomorphisms on base space. This allows to interpret the 
BPS Skyrme matter as an incompressible liquid. Moreover, the solitons of this 
theory are of compact type which results in a finite range of interactions 
(contact-like interactions). Therefore, the BPS Skyrme model cures the above 
mentioned problems of the standard Skyrme model, at least on a qualitative
level, although it does so in a rather 
radical way. In addition, these properties are $N_c$ independent. The zero 
binding energy, liquid state of matter and the contact interaction occur
for all $N_c$, no matter how $N_c$ enters into the parameters of the BPS 
model.  As a consequence, the BPS Skyrme model apparently is a good guess 
for certain aspects of a low 
energy effective action for $N_c=\infty$, although there does not seem to
exist yet 
an obvious large $N_c$ limit based directly on QCD which would produce
just the BPS Skyrme model as its leading order.
\\
We want to remark that there exists another BPS generalization 
of the Skyrme model recently proposed by Sutcliffe \cite{sutcliffe bps} based 
on dimensional reduction of the (4+1) dimensional 
Yang-Mills (YM) theory, where the $SU(2)$ 
Skyrme field is accompanied by an infinite tower of the vector and tensor 
mesons. It also gives the linear energy-charge relation, as an inherited 
property from the pertinent self-dual sector of the YM theory, 
and potentially zero 
binding energy. Whether the large $N_c$ problems of the standard Skyrme model
can be cured in this model, and whether the resulting nuclear matter is in a
cristal or liquid state, still has to be investigated. Also the addition of a
potential seems to be difficult if one wants to maintain the relation with 
4+1 Yang-Mills theory.
%%%%%%%%%%%%%%%%%%%%%%%%%%%%%%%%%%%%%%%%%%%%%%%%%%%%%%%%%%%%%%%%%%
\section{The BPS Skyrme models}
%%%%%%%%%%%%%%%%%%%%%%%%%%%%%%%%%%%%%%%%%%%%%%%%%%%%%%%%%%%%%
Let us, then, consider the proposed family of models, which we will call 
{\it the BPS Skyrme models} 
 \begin{equation}
L_{06}=\frac{\lambda^2}{24^2 } \;\left[  \mbox{Tr} \;  ( \epsilon^{\mu 
\nu \rho \sigma} U^{\dagger} \partial_{\mu} U \;
U^{\dagger} \partial_{\nu} U \;
U^{\dagger} \partial_{\rho} U) \right]^2 - \mu^2 V(U,U^{\dagger}).
\end{equation}
The subindex 06 refers to the fact that in the above Lagrangian only a
potential term without derivatives and a term sextic in 
derivatives are present.\footnote{We remark that models which are 
similar in some aspects, although with a 
different target space geometry,  have been
studied in \cite{Tchr 1}, \cite{Tchr 2}.
Further, the model studied here and in the letter \cite{proposal}, 
as well as its "baby Skyrme" 
version in 2+1 dimensions have already been introduced in  \cite{Tchr 3}, 
where
the main aim was to study more general properties of Skyrme models in 
any dimension. For further discussion of the (2+1) dimensional version of 
it see \cite{GP}, \cite{baby}, \cite{FN}, and, from a more
geometric point of view, \cite{speight2}.
} The sextic part of the action is nothing but the topological current 
density squared.  
Here, we use the standard parametrization by means of a real scalar $\xi$ 
and a three component unit vector $\vec{n}$ field 
($\vec \tau$ are the Pauli matrices), 
$$
U=e^{i \xi \vec{n} \cdot \vec{\tau}}.
$$
The vector field may be related to a complex scalar $u$ by the 
stereographic projection
$$
\vec{n}=\frac{1}{1+|u|^2} \left( u+\bar{u}, -i ( u-\bar{u}),
|u|^2-1 \right).
$$
Assuming for the potential
$$
V=V({\rm tr} (U+U^\dagger) )=V(\xi)
$$
(which we assume for the rest of the paper) we get      
\begin{equation}
L_{06}= -\frac{  \lambda^2 \sin^4 \xi}{(1+|u|^2)^4} \;\left(  
\epsilon^{\mu \nu \rho \sigma} \xi_{\nu} u_{\rho} \bar{u}_{\sigma} \right)^2
-\mu^2 V(\xi)\end{equation}
The pertinent equations of motion read
$$ \frac{\lambda^2 \sin^2 \xi}{(1+|u|^2)^4} \partial_{\mu} ( \sin^2 \xi \; 
H^{\mu}) + \mu^2 V'_{\xi}=0,$$
$$ \partial_{\mu} \left( \frac{K^{\mu}}{(1+|u|^2)^2} \right)=0,$$
where $$ H_{\mu} = \frac{\partial (  \epsilon^{\alpha \nu \rho \sigma} 
\xi_{\nu} u_{\rho} \bar{u}_{\sigma})^2}{ \partial \xi^{\mu}}, \;\;\; K_{\mu} = 
\frac{\partial (  \epsilon^{\alpha \nu \rho \sigma} \xi_{\nu} u_{\rho} \bar{u}
_{\sigma})^2}{\partial \bar{u}^{\mu}}.$$
These objects obey the useful formulas 
$$H_{\mu} u^{\mu}=H_{\mu} \bar{u}^{\mu}=0, 
\; K_{\mu}\xi^{\mu}=K_{\mu} u^{\mu}=0, \;\; H_{\mu} \xi^{\mu}=K_{\mu} 
\bar{u}^{\mu} = 2 (  \epsilon^{\alpha \nu \rho \sigma} \xi_{\nu} u_{\rho} 
\bar{u}_{\sigma})^2. $$ 
%%%%%%%%%%%%%%%%
\subsection{Symmetries}
%%%%%%%%%%%%%%%
Apart from the standard Poincare symmetries, the model has an infinite 
number of
target space symmetries. The sextic term alone is the square of the pullback 
of the volume form on the target space $S^3$, where this target space volume
form reads explicitly
\begin{equation}
dV= -i\frac{\sin^2 \xi}{(1+|u|^2)^2}d\xi du d\bar u
\end{equation}
and the exterior (wedge) product of the differentials is understood.  
Therefore, the sextic term alone is invariant under all target space
diffeomorphisms which leave this volume form invariant (the volume-preserving
diffeomorphisms on the target $S^3$). The potential term in general does not
respect all these symmetries, but depending on the specific choice, it may
respect a certain subgroup of these diffeomorphisms. Concretely, for
$V=V(\xi)$, the potential is invariant under those volume-preserving 
target space diffeomorphisms which do not change $\xi$, that is, which act
nontrivially only on $u, \bar u$. Since $u$ spans a two-sphere in target space,
these transformations form a one-parameter family of the groups of the 
area-preserving
diffeomorphisms on the corresponding target space $S^2$ (one-parameter family
because the transformations may still depend on $\xi$, although they act
nontrivially only on $u, \bar u$). Both the Poincare
transformations and this family of
area-preserving target space diffeomorphisms are
symmetries of the full action, so they are Noether symmetries with the
corresponding conserved currents. The latter symmetries may, in fact, be
expressed in terms of the generalized integrability, as we briefly discuss in
the next subsection. 
\\
The energy functional for static fields has an additional group of infinitely
many symmetry transformations, as we want to discuss now. These symmetries are
not symmetries of the full action, so they are not of the Noether type, but
nevertheless they are very interesting from a physical point of view, as we
will see in the sequel. The energy functional for static fields reads
\begin{equation}
E=\int d^3 x \left(  \frac{\lambda^2 \sin^4 \xi}{(1+|u|^2)^4} 
(\epsilon^{mnl} i \xi_m u_n\bar{u}_l)^2 +\mu^2 V \right)
\end{equation}
and we observe that both $d^3 x$ and $\epsilon^{mnl} i \xi_m u_n\bar{u}_l$ are
invariant under coordinate transformations of the base space coordinates $x_j$
which leave the volume form $d^3 x$ invariant. So this energy functional has
the volume-preserving diffeomorphisms on base space as symmetries. These
symmetries are precisely the symmetries of an incompressible ideal fluid,
which makes them especially interesting in the context of applications to
nuclear matter. Indeed, the resulting field theory is able to reproduce some
basic features of the liquid droplet model of nuclei, see
e.g. \cite{proposal}, \cite{slobo} and the discussion below. 
%%%%%%%%%%%%%%%%%%%%%%%%%%%%%%%%%%%%%%%%%%
\subsection{Integrability}
%%%%%%%%%%%%%%%%%%%%%%%%%%%%%%%%%%%%%%%%%%
The BPS Skyrme model is integrable in the sense that there 
are infinitely 
many conserved charges. Indeed, it belongs to a family of models integrable 
in the sense of the generalized integrability \cite{alvarez},
\cite{pullback}. 
To show that we introduce 
$$ \mathcal{K}^{\mu}=\frac{K^{\mu}}{(1+|u|^2)^2}.$$
The currents are
$$ J_{\mu}=\frac{\delta G}{\delta \bar{u}} \bar{\mathcal{K}}^{\mu} - 
\frac{\delta G}{\delta u} \mathcal{K}^{\mu}, \;\;\; G=G(u,\bar{u},\xi) $$
where $G(u,\bar u,\xi)$ is an arbitrary real function of its arguments.
Then, 
$$\partial^{\mu} J_{\mu}= G_{\bar{u}\bar{u}} \bar{u}_{\mu} 
\bar{\mathcal{K}}^{\mu} + G_{\bar{u}u} u_{\mu} \bar{\mathcal{K}}^{\mu} + 
G_{\bar{u}} \partial_{\mu} \bar{\mathcal{K}}^{\mu} -G_{u \bar{u}}
\bar{u}_{\mu} 
\mathcal{K}^{\mu} -G_{uu} u_{\mu} \mathcal{K}^{\mu} - G_{u} \partial_{\mu} 
\mathcal{K}^{\mu}$$ $$ + G_{\bar{u} \xi} \xi_{\mu} \bar{\mathcal{K}}^{\mu} - 
G_{u \xi} \xi_{\mu} \mathcal{K}^{\mu}=0, $$
where we used that $u_{\mu} \mathcal{K}^{\mu}=\xi_{\mu} \mathcal{K}^{\mu}=
0 $, $ \bar{u}_{\mu} \mathcal{K}^{\mu}=u_{\mu} \bar{\mathcal{K}}^{\mu},$ which 
follow from the previous identities. Finally using the field equations for 
the complex field, 
one, indeed, finds an infinite number of conserved currents.
These currents are a higher dimensional generalization of 
those constructed for the pure baby Skyrme model \cite{integr} and are 
generated by the relevant subgroup of the 
volume-preserving diffeomorphisms on the target space as discussed in the
previous subsection, see also
\cite{ab-dif}. 
\\
We remark that the existence of infinitely many conservation
laws (integrability) together with the possibility to reduce the static field
equations to a system of ordinary differential equations via the right ansatz
(the hedgehog ansatz in our case) leads to the existence of infinitely many
exact solutions, as we shall see in next subsections. This observation lends 
further
credibility to the conjecture that the relation between integrability (in the
sense of infinitely many conservation laws) and solvability 
is always true in higher-dimensional
field theories. The conjecture holds true in all concrete cases we are aware
of. Nevertheless, a deeper mathematical understanding or even a proof, which
would advance our understanding of integrability in higher dimensions, 
is still under investigation.

%%%%%%%%%%%%%%%%%%%%%%%%%%%%%%%%%%%%%%%%%%
\subsection{Bogomolny bound}
%%%%%%%%%%%%%%%%%%%%%%%%%%%%%%%%%%%%%%%%%%
In the BPS Skyrme model, there exists the following 
Bogomolny bound for the energies $E$ of
static solutions
\begin{equation}
E \ge 2\lambda \mu \pi^2 C[V] |B|
\end{equation}
where $B$ is the baryon number (topological charge). Further, $C[V]$ is a
calculable number which depends on the potential but {\em not} on the specific
solution. Therefore, for a given theory (i.e., a given potential), the bound is, indeed, topological. 
\\
For a proof, we write the energy functional as
$$
E=\int d^3 x \left(  \frac{\lambda^2 \sin^4 \xi}{(1+|u|^2)^4} 
(\epsilon^{mnl} i \xi_m u_n\bar{u}_l)^2 +\mu^2 V \right) = 
$$ 
$$ 
= \int d^3 x \left( \frac{\lambda \sin^2 \xi}{(1+|u|^2)^2} 
\epsilon^{mnl} i \xi_mu_n\bar{u}_l \pm \mu \sqrt{V} \right)^2 \mp \int d^3 x 
\frac{2\mu \lambda \sin^2 \xi \sqrt{V}}{(1+|u|^2)^2} \epsilon^{mnl} i \xi_m 
u_n \bar{u}_l 
$$
$$ 
\geq  \mp \int d^3 x \frac{2\mu \lambda \sin^2 \xi \sqrt{V}}{(1+|u|^2)^2} 
\epsilon^{mnl} i \xi_m u_n \bar{u}_l =
$$
\begin{equation} \label{bobo}
\pm (2\lambda \mu \pi^2 )\left[ \frac{-i}{\pi^2}
\int d^3 x \frac{ \sin^2 \xi \sqrt{V}}{(1+|u|^2)^2} 
\epsilon^{mnl}  \xi_m u_n \bar{u}_l \right] 
\equiv 2\lambda \mu \pi^2 C[V] |B|
\end{equation}
where the sign has to
be chosen appropriately (upper sign for $B>0$). 
If we replace $V$ by one in the last expression in brackets, and $C[V]$ by one, then this expression is just the topological charge, so its topological nature is obvious. Equivalently, this expression is just the
base space integral of the pullback of the volume form on the
target space $S^3$, normalized to one, and this second interpretation may be easily generalized to nontrivial $V$. Indeed, we just have to introduce
a new target space coordinate $\bar\xi $ such that
\begin{equation} \label{xi-xi'}
\sin^2 \xi \sqrt{V(\xi)} \, d\xi = C[V]  \sin^2 \bar \xi  \, d \bar\xi  .
\end{equation}
The constant $C[V]$ and a second constant $C'$, which is provided by
the integration of Eq. (\ref{xi-xi'}), are needed to impose the two conditions
$\bar\xi (\xi=0)=0$ and $\bar \xi (\xi =\pi) =\pi$, which have to hold if 
$\bar\xi$ is a
good coordinate on the target space $S^3$. Obviously, $C[V]$ depends on the
potential $V(\xi)$.
E.g.,  for the standard Skyrme potential $V=1-\cos\xi$, $C[V]$ is
$$
C[V] = \frac{32\sqrt{2}}{15\pi} .
$$
The corresponding Bogomolny (first order) equation is
\begin{equation}
\frac{\lambda \sin^2 \xi}{(1+|u|^2)^2} \epsilon^{mnl} i \xi_mu_n\bar{u}_l 
= \mp \mu \sqrt{V} 
\end{equation}
and is satisfied by all soliton solutions which we shall encounter in this
article.
This proof for the Bogomolmy bound has been given already in \cite{proposal} and is repeated here for the convenience of the reader
(the proof for the analogous 2+1-dimensional theory has been given in
\cite{ward}, \cite{baby}, \cite{speight2}).
%%%%%%%%%%%%%%%%%%%%%%%%%%%%%%%%%%%%%%%%%%
\subsection{Exact solutions}
%%%%%%%%%%%%%%%%%%%%%%%%%%%%%%%%%%%%%%%%%%
We are interested in static topologically nontrivial solutions. Thus $u$ must 
cover the whole complex plane ($\vec{n}$ covers at least once $S^2$) 
and $\xi \in [0,\pi]$. The natural (hedgehog) ansatz is
\begin{equation}
\xi = \xi (r), \;\;\; u(\theta, \phi) = g (\theta) e^{in \phi}.
\end{equation}
Then, the field equation for $u$ reads
$$ 
\frac{1}{\sin \theta} \partial_{\theta} \left( \frac{ g^2g_\theta}{(1+g^2)^2 
\sin \theta} \right) - \frac{gg_\theta^2}{(1+g^2)^2\sin^2 \theta}=0,
$$
and the solution with the right boundary condition is
\begin{equation} \label{u-sol}
g(\theta) = \tan \frac{\theta}{2}.
\end{equation}
Observe that this solution holds for all values of $n$. 
The equation for the real scalar field is 
$$
\frac{n^2\lambda^2 \sin^2 \xi }{2r^2} \partial_r \left(\frac{\sin^2 \xi \; 
\xi_r}{r^2} \right) - \mu^2 V_{\xi}=0.
$$ 
This equation can be simplified by introducing the new variable 
\begin{equation}
z=\frac{\sqrt{2}\mu r^3}{3 |n|\lambda} .
\end{equation} 
It reads
\begin{equation} \label{xi-eq}
\sin^2 \xi \; \partial_z \left(\sin^2 \xi \; \xi_z\right) -  V_{\xi}=0,
\end{equation}
and may be integrated to 
\begin{equation} 
 \frac{1}{2} \sin^4 \xi \; \xi^2_z=V, \label{bps eq}
\end{equation}
where we chose a vanishing integration constant to get finite energy solutions. 
We remark that this first integration of the field equation is equivalent to
a Bogomolny equation and, thus, to a Bogomolny bound for the dimensionally
reduced, effectively one-dimensional problem. It can be proved that also for
the full theory, without any symmetry reduction, there exists a Bogomolny
bound and a Bogomolny equation which is obeyed by all the solutions
we find in the sequel. 
In terms of the variable $r$ the integrated (Bogomolny) equation reads
\begin{equation} \label{bogo-r}
\frac{n^2 \lambda^2}{4\mu^2 r^4} \sin^4 \xi \xi_r^2 =V
\end{equation}
and it is this form which will be useful for the discussion of the energy to
be performed next. Indeed, the energy is
\begin{equation}
E=\int d^3x \left( -\frac{\lambda^2 \sin^4 \xi}{(1+|u|^2)^4} 
(\nabla_r \xi )^2 ( \nabla_{\theta} u \nabla_{\phi} \bar{u} - \nabla_{\phi} u 
\nabla_{\theta} \bar{u})^2 +\mu^2 V  \right).
\end{equation}
or, after inserting the hedgehog ansatz with the solution (\ref{u-sol}) for
$u$, 
\begin{equation}
E = 4\pi \int r^2 dr \left( \frac{\lambda^2 n^2\sin^4 \xi}{4r^4} \xi^2_r 
+\mu^2 V  \right) .
\end{equation}
It follows from the Bogomolny equation for $r$, (\ref{bogo-r}), that the
sextic term and the potential contribute the same amount to the energy density
for arbitrary values of $r$. Therefore, we may further simplify the expression
for the energy like
\begin{equation}
E = 4\pi \cdot 2\mu^2 \int r^2 dr V(\xi (r))= 4 \sqrt{2}\pi 
\mu \lambda |n| \int dz V (\xi (z))  .
\end{equation}
Further, we may already draw some qualitative conclusions about the behaviour
of the energy density profiles for different types of potentials. Finiteness
of the energy requires that the fields take values in the vacuum manifold of
the potential $V$ in the limit $r\to \infty$. For the class of potentials
$V=V(\xi)$ we consider this just means that $\lim_{r\to\infty}V(\xi(r))=0$.
Further, the topology of skyrmion fields requires that the matrix field $U$
takes a constant, direction-independent value in the limit $r\to
\infty$. Within the hedgehog ansatz this implies that the field $\xi$ must
take one of its two boundary values $\xi =0,\pi$ in this limit. For skyrmions
with finite energy, therefore, at least one of these two boundary values must
belong to the vacuum manifold of the potential. Without loss of generality,
let us assume that $\xi$ takes the value $\xi =0$ in the limit $r\to\infty$. 
For a wide class of potentials
this implies that $\xi$ must take the opposite boundary value $\xi =\pi$ at
$r=0$, because it follows easily from (\ref{bogo-r}) that $\xi$ is a
monotonous function of $r$ in the region where $V\not= 0$. 
These observations lead to the following
conclusions. For one-vacuum potentials with the only vacuum at $\xi=0$, the
energy density cannot be zero inside the skyrmion. If, in addition, the
potential is a monotonous function of $\xi$ in the range of $\xi$, then the
energy density is a monotonous function of $r$ and takes its maximum value at
$r=0$, i.e., the soliton is of the core type. If the potential has the two
vacua $\xi =0,\pi$, then the energy density is zero also at $r=0$, and the
soliton is of the shell type. For more complicated vacuum manifolds of $V$,
more complicated soliton structures emerge, but they may still be found by a
variant of the simple qualitative reasoning applied in this paragraph. 
We remark that a qualitatively
similar relation between the vacuum manifold of the potential and the skyrmion
structure also is observed in the original Skyrme model with a potential. The
difference is that in the latter case this relation is the result of
complicated, three-dimensional numerical integrations, whereas in our case it
follows from some simple, analytical arguments.   
%%%%%%%%%%%%%%%%%%%%%%%%%%%%%%%%%%%%%%%
\subsection{The Skyrme potential}
%%%%%%%%%%%%%%%%%%%%%%%%%%%%%%%%%%%%%%%%%%
The first obvious possibility is to consider the standard Skyrme potential 
\begin{equation}
V=\frac{1}{2}\mbox{Tr} (1-U) \;\; \rightarrow \;\; V(\xi)=1- \cos \xi. 
\end{equation}
Imposing the boundary conditions for
topologically non-trivial solutions we get 
\begin{equation}
\xi = \left\{
\begin{array}{lc}
2 \arccos \sqrt[3]{ \frac{3z}{4} } & z \in \left[0,\frac{4}{3} \right] \\
0 & z \geq \frac{4}{3}.
\end{array} \right. \label{xi sol}
\end{equation}
The corresponding energy is
\begin{eqnarray}
E &=&  8 \sqrt{2}\pi \mu \lambda |n| \int_0^{4/3} 
\left(1-\left( \frac{3z}{4} \right)^\frac{2}{3} \right)dz = 
\frac{64\sqrt{2} \pi}{15} \mu \lambda |n| . \label{E-skyrmepot}
\end{eqnarray}
 The solution is of 
the compacton type, i.e., it has a finite support \cite{comp}
(compact solutions of a similar
type in different versions of the
baby Skyrme models have been found in \cite{GP},
\cite{comp-bS}).
The function $\xi$ is continuous 
but its first derivative is not. The jump of the derivative is, in fact, 
infinite at the compacton boundary
$z=4/3$, as the left derivative at this point tends to minus infinity. 
Nevertheless, the energy density and the topological charge density
(baryon number density) are continuous functions at the compacton boundary,
and the field equation (\ref{xi-eq}) is well-defined there.
The reason is that $\xi_z$ always appears in the
combination $\sin^2 \xi \, \xi_z$, and this expression is finite (in fact, zero)
at the compacton boundary. We could make the discontinuity disappear
altogether by introducing a new variable $\tilde \xi$ instead of $\xi$
which satisfies
$$
\tilde \xi_z = \sin^2 \xi \, \xi_z .
$$
We prefer to work with $\xi$ just because this is the standard
variable in the Skyrme model.  
\\
In order to extract the energy density it is useful to rewrite the energy with
the help of the rescaled radial coordinate
\begin{equation}
\tilde r = \left( \frac{\sqrt{2}\mu}{4  \lambda} \right)^\frac{1}{3} r 
\equiv \frac{r}{R_0} =
\left(\frac{3|n|z}{4}\right)^\frac{1}{3}
\end{equation}
(here $R_0$ is the compacton radius)
like
$$
E = 8 \sqrt{2}  \mu \lambda \left( 4 \pi \int_0^{|n|^\frac{1}{3}} d\tilde r 
\tilde r^2 (1- |n|^{-\frac{2}{3}}\tilde r^2) \right) 
$$
such that the energy density per unit volume (with the unit of length set by
$\tilde r$) is
\begin{eqnarray}
{\cal E}&=&   8 \sqrt{2} \mu \lambda (1- |n|^{-\frac{2}{3}} \tilde r^2 ) 
\quad \mbox{for} \quad 0\le \tilde r \le |n|^\frac{1}{3} \nonumber \\
&=& 0\quad \mbox{for} \quad \tilde r > |n|^\frac{1}{3}. 
\end{eqnarray}
$\tilde r$ does not depend on the topological charge $B=n$, so the dependence
of ${\cal E}$ on $n$ is explicit.
\\
\begin{figure}[h!]
\includegraphics[angle=0,width=0.55 \textwidth]{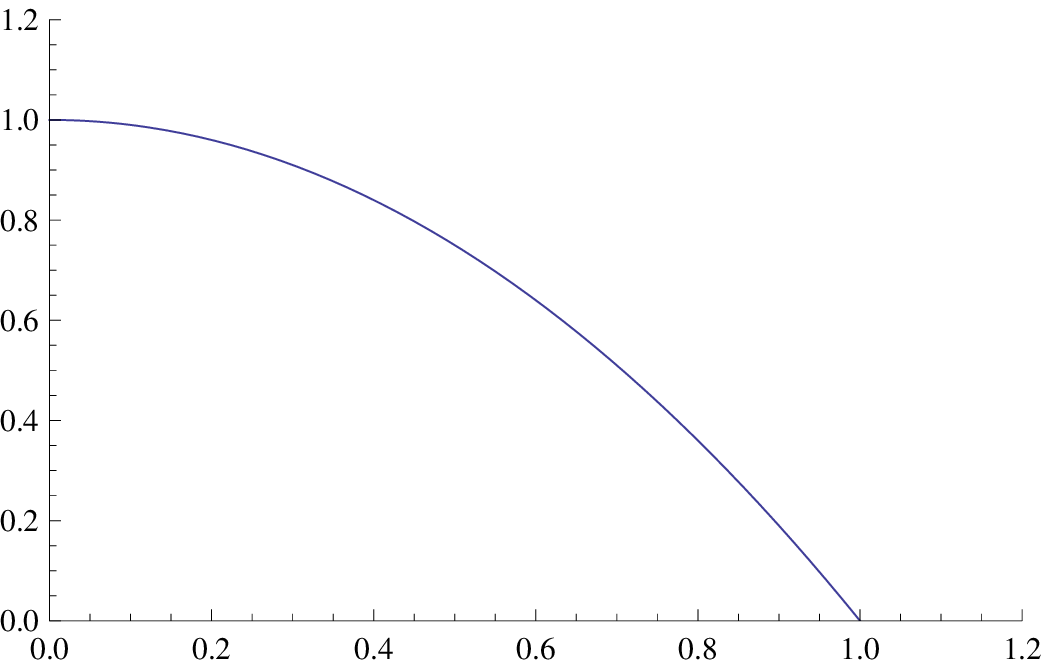}
\includegraphics[angle=0,width=0.55 \textwidth]{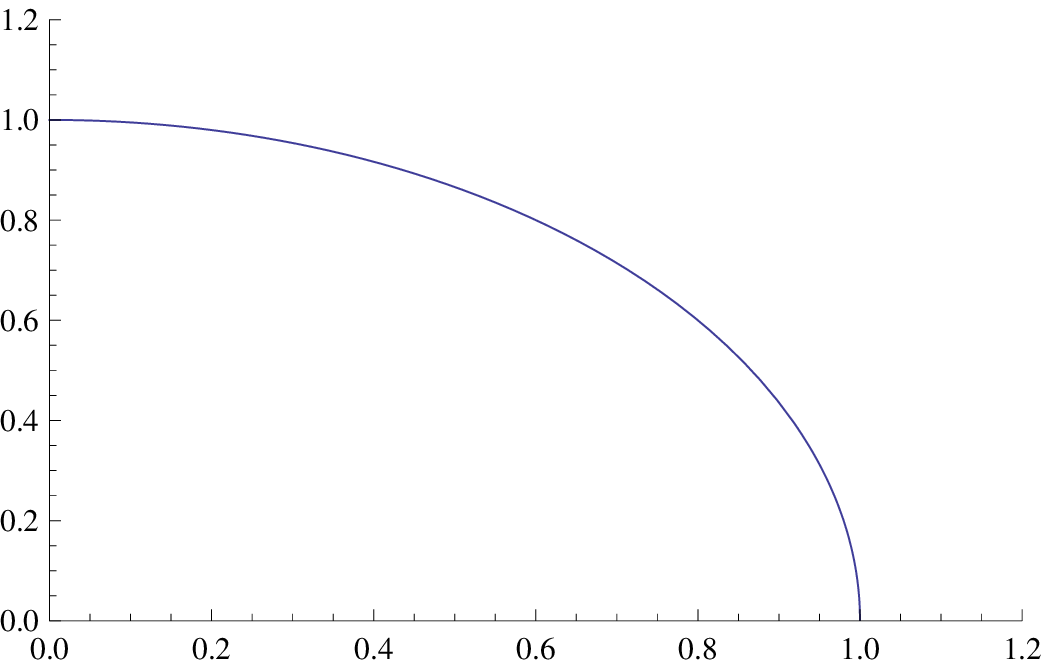}
\caption{Normalized energy density (left figure) and topological charge 
density (right figure) as a function of the
rescaled radius $\tilde r$, for topological charge n=1. For $|n|>1$, the
height of the densities remains the same, whereas their radius grows like
$|n|^\frac{1}{3}$  }\label{rys1}
\end{figure}
In the same fashion we get for the topological charge (baryon number), see
e.g. chapter 1.4 of \cite{mak}
\begin{eqnarray} \label{Bcharge}
B &=& -\frac{1}{\pi^2} \int d^3 x \frac{\sin^2 \xi }{(1+|u|^2)^2}
i\epsilon^{mnl} \xi_m u_n \bar u_l 
= - \frac{2n}{\pi} \int dr \sin^2 \xi \, \xi_r  \\
&=& - \frac{2n}{\pi} \int dz \sin^2 \xi \, \xi_z =
\frac{4n}{\pi} \int_0^\frac{4}{3} dz
\left( 1-\left(\frac{3}{4}\right)^\frac{2}{3} z^\frac{2}{3}\right)^\frac{1}{2}
\nonumber \\
&=& \mbox{sign} (n) \frac{4}{\pi^2} \left(4\pi \int_0^{|n|^\frac{1}{3}} 
d\tilde r \tilde
r^2 (1- |n|^{-\frac{2}{3}}\tilde r^2)^\frac{1}{2} \right) =n \nonumber
\end{eqnarray}
and for the topological charge density per unit volume
\begin{eqnarray}
{\cal B} &=&  \mbox{sign} (n) \frac{4}{\pi^2} (1- |n|^{-\frac{2}{3}}\tilde
r^2)^\frac{1}{2} \quad \mbox{for} \quad 0\le \tilde r \le 
|n|^\frac{1}{3} \nonumber \\
&=& 0\quad \mbox{for} \quad \tilde r > |n|^\frac{1}{3}. 
\end{eqnarray}
Both densities are zero outside the compacton radius $\tilde r
=|n|^\frac{1}{3}$. 
We remark that the values of the densities at the center $\tilde r=0$ are
independent of the topological charge $B=n$, whereas the radii grow like
$n^\frac{1}{3}$.  For $n=1$, we plot the two densities in Fig. \ref{rys1},
where we normalize both densities (i.e., multiply them by a constant) such
that their value at the center is one.
%%%%%%%%%%%%%%%%%%%%%%%%%%%%%%%%%%%%%%
\section{Some phenomenology of nuclei}
%%%%%%%%%%%%%%%%%%%%%%%%%%%%%%%%%%%%%%
After having discussed the properties of the Lagrangian and its classical
solutions in the preceding sections, let us now try to apply it to the
description of some properties of nuclei. After all, this possibility is one
of the rationales for the original Skyrme model and its generalizations.
We do, of course,  not consider this model by itself as the correct 
effective model of 
QCD, but we want to see how far solitons of this integrable model can be
related in one way or another to some properties of real baryons.
Here we shall first focus on the classical theory and solutions, and we
will find that at this level the model already reproduces quite well some
properties of the nuclear drop model. 
In a next step, we perform the
semi-classical quantization of the (iso)-rotational degrees of freedom of the
$B=1$ soliton, i.e., the nucleon.
Further, we choose the standard Skyrme potential $V=1-\cos \xi$ for simplicity
throughout this section.
%%%%%%%%%%%%%%%%%%%%%%%%%%%%%%%%%%%%%%
\subsection{Classical aspects}
%%%%%%%%%%%%%%%%%%%%%%%%%%%%%%%%%%%%%%
We find immediately that the classical solutions of
the BPS Skyrme model seem to describe surprisingly well some static 
properties of nuclei. As was discussed already in \cite{proposal}, it 
provides an alternative starting point for an effective soliton model of 
baryons, which by construction is much more topological in nature. 
Let us present and further elaborate on these results.

\vspace*{0.1cm}
{\it Mass spectrum and linear energy-charge relation.} As a consequence of the
BPS nature of the classical solutions, 
the energy of the solitons is proportional to the topological (baryon) 
charge
$$ E=E_0 |B|, $$
where $E_0=64\sqrt{2}\pi \mu\lambda / 15$. Such a linear dependence is a 
well established fact in nuclear physics.  For the moment (i.e., in the
context of the purely classical reasoning), let us fix the energy 
scale by assuming that $E_0 = 931.75 $ MeV. This is equivalent to the
assumption  
that the mass of the solution with $B=4$ is equal to the mass 
of He$^4$. One usually assumes this value because  
the ground state of 
He$^4$ has zero spin and isospin \cite{massive2}. Therefore, possible 
corrections to the mass from spin-isospin interactions are absent. In 
Table \ref{table} we compare energies of the solitons in the BPS model 
with experimental values and energies obtained in the vector-Skyrme 
\cite{vec skyrme} and standard massive Skyrme model \cite{massive1} 
(We use the numerical data, if accessible, or calculate them from fitted 
functions \cite{massive1}. The energy scale is set by the same prescription). 
It is interesting to note that instead of the approximate $7\%$ 
accuracy typical for the soliton energies of standard Skyrme theories
we get maximally only  a $0.7\%$ discrepancy. 
Besides, the masses of the BPS Skyrme model solitons are
slightly smaller than the experimental masses in {  \em all} cases (except for
the He$^4$ used for the fit, of course). This goes into the right direction,
because the (iso-) rotational excitation energies should be added to the
classical soliton masses (except for the He$^4$, of course)
for a more reliable comparison with physical masses of nuclei.   

\vspace*{0.1cm}
{\it No binding energy.} It follows from the BPS nature of the model that the 
binding energy is zero. This is different from the standard Skyrme model, 
where binding energies are rather big. For example, the energy of the baryon 
number two skyrmion exceeds the topological energy bound by $23 \%$. Of 
course, such binding energies are significantly larger than experimentally 
observed, which usually do not reach $1\%$. Therefore, as pointed out by P. 
Sutcliffe \cite{sutcliffe1}, a BPS Skyrme theory seems to be a better 
starting point to get realistic binding energies. Small (non-zero) binding 
energies could be produced by  small perturbations around a BPS theory. 

\vspace*{0.1cm}
{\it Size of nuclei and compactons.} Due to the compact nature of the
solitons, their radius is well defined 
and can be easily computed $$ 
R_B= R_0 \sqrt[3]{B}, \;\;\; R_0=\left( \frac{2\sqrt{2} \lambda }{\mu} 
\right)^{\frac{1}{3}},$$
which again reproduces the well-known experimental
relation. The numerical value which best reproduces the known radii of nuclei
is approximately  $R_0=1.25$ fm.  
\\
Further, the compact nature of our solutions 
probably can be viewed as an 
advantage of the model rather than a defect. In fact, the absence of 
interactions (or, more precisely, of  {\it finite range interactions}) 
corresponds quite well with the very short range of forces between nuclei. 

\vspace*{0.1cm}
{\it Core type energy density.} For the solution of Section 2.5, 
which provides spherically symmetric energy densities
for all baryon numbers, the resulting energy density 
takes its
maximum value at the 
origin.  
It is of some interest to compare this result with the densities in the 
standard 
massless or massive Skyrme models. For the massless Skyrme model, solitons are 
geometrically complicated shell-like structures with empty space regions 
inside \cite{massless}. In addition, the size of the shell-skyrmions grows 
like $\sqrt{B}$, which is in contradiction to the experimental 
data. In the case of the massive Skyrme model, the situation is slightly more 
subtle \cite{massive1}, \cite{massive2}, \cite{sutcliffe1}. 
The proper size-charge relation has 
been reported \cite{massive1}. Moreover, depending on the mass of the pion 
field and baryon number, squeezed clustered solutions, instead of shell 
ones, begin to be preferred. Precisely speaking, for a fixed value of the 
mass parameter, the first few skyrmions possess a shell-like structure, whereas 
for higher baryon charge a clustered solution is the true minimum. The 
critical charge, for which a shell-skyrmion occurs, seems to be smaller if  
the mass is increased \cite{massive2}. However, even for the physically 
acceptable 
value $m=1$ (which is more or less twice the bare pion mass), skyrmions with 
$B \leq 9$ are shells. In the modern interpretation this problem can be 
cured by treating the massive parameter as a renormalized pion mass which 
should be adjusted to best reproduce observed data \cite{massive2}, 
\cite{renorm mass}.  Then, increasing $m$ one gets rid of unwanted shell 
solutions, leaving only clustering ones. 
\\
\begin{table}
\begin{center}
\begin{tabular}{|c|c|c|c|c|}
\hline
B & $E_{experiment}$ & $E_{BPS}$ & $E_{vec \; \; Skyrme}$ &  $E_{Skyrme}$ \\
\hline 
1 & 939 &  931.75 & 996  &1024 \\
2 & 1876 &  1863.5 &  1999 & 1937 \\
3 & 2809 &  2795.25  &  2913 & 2836 \\
4 & 3727 &  3727 &  3727 & 3727  \\
6 & 5601 &  5590.5 & - & 5520  \\
8 & 7455 &  7454 & - & 7327 \\
10 & 9327 &  9317.5 & - & 9113 \\
\hline
\end{tabular}  
\caption{Energies of the solutions in the pure Skyrme model, compared with 
masses for the vector-Skyrme and Skyrme model, as well as with the 
experimental date. All numbers are in MeV} \label{table}
\end{center}
\end{table}
Let us also notice that there is a reminiscence of this clustering phenomenon 
in the BPS Skyrme model, even though it is quite trivial. Namely, due to 
the compact and BPS nature of the solutions of the BPS Skyrme model, it is 
possible to construct a collection of separated components provided they are 
sufficiently separated (they do not touch each other). Such a clustered 
configuration has a total baryon number equal to the sum of the components. 
In the BPS Skyrme model, none of these clustered (multi-center) solutions 
is energetically preferred, which again is a simple outcome of the BPS origin 
of the solitons.    
\\
Finally, the values of the energy and charge densities of the solutions of
Section 2.5 at the center do not depend on the baryon number, which, again, is
a property which holds reasonably well for physical nuclei.

\vspace*{0.1cm}
{\it The liquid drop property.}  
The energy functional for static field configurations has
the volume-preserving diffeomorphisms on the
three-dimensional base space as symmetries. In
physical terms, all deformations of solitons
which correspond to these volume-preserving
diffeomorphisms may be performed without any cost in energy. 
But these deformations are exactly
the allowed deformations for an ideal, incompressible droplet of liquid when
surface contributions to the energy are neglected. 
These symmetries are not symmetries of a physical nucleus. A physical nucleus
has a definite shape, and deformations which change this shape cost energy.
Nevertheless, deformations which respect the local volume conservation (i.e.,
deformations of an ideal incompressible liquid) cost much less energy
than volume-changing deformations, as an immediante consequence 
of the liquid drop model of nuclear matter. 
This last observation also further explains the nature of the approximation our
model provides for physical nuclei. It reproduces some of the classical
features of the
nuclear liquid drop model at least on a qualitative level, and the huge
amount of symmetries of the model is crucial for this fact.  
Its soliton energies, e.g., correspond to the bulk (volume) contribution of
the liquid drop model, with the additional feature that the energies are
quantized in terms of a topological charge.   
\\
This should be contrasted with the expected behaviour for large baryon number
for the standard Skyrme model. In the standard Skyrme model, there remain some
long range forces between different Skyrmions, 
whose attractive or repulsive character depends on the
relative orientation of the Skyrmions. As a consequence, it is expected that
for large baryon number the energy-minimizing configurations are Skyrmion
crystals, where all the Skyrmions are brought into the right positions and
orientations to minimize the total energy. For physical nuclear matter, there
is no sign of this crystal type behaviour. Instead, nuclear matter seems to be
in a liquid state, which is well described by our BPS Skyrme model.  
\\
Let us remark that in QCD at 
$N_c=\infty$ the instanton liquid becomes incompressible, as well
\cite{nowak}.
Whether this appearence of an incompressible liquid at large $N_c$
both in the BPS Skyrme model and in the instanton liquid is more than a 
mere coincidence remains to be seen.
\\
Let us also remark that a liquid drop like behaviour for nuclear matter has
been established recently in a rather different approach. Indeed, 
in \cite{forcrand} the authors
obtain both the bulk (volume) term and the surface
contribution of the liquid drop model to the nuclear masses
from an ab initio very idealized and simplified QCD lattice
computation, with massless quarks and infinite coupling. In addition, they
also find, as in our case, the absence of
pseudoscalar meson exchange forces (see next paragraph).

\vspace*{0.1cm}
{\it Absence of pion fluctuations.} 
In the model, both the quadratic and the quartic kinetic terms are absent. As
a consequence, neither propagating pions nor the two-body interaction between
pions can  be described in the model. Nevertheless, already at the classical
level the model seems to describe some nuclear properties reasonably well,
which seems to indicate that in certain
circumstances the sextic term could be more important than the terms $L_2$ and
$L_4$. The quadratic term is kinetic in nature, whereas the
quartic term provides, as a leading behaviour, two-body interactions. On the
other hand, the sextic term is essentially topological in nature, being the
square of the topological current (baryon current). So in circumstances where
our model is successful this seems to indicate that a {\em collective} 
(topological)
contribution to the nucleus is more important than kinetic or two-body
interaction contributions. This behaviour is, in fact, not so surprising for a
system at strong coupling (or for a strongly non-linear system).  
A first consequence of the absence of pions is the compact nature of the
solutions, i.e., the absence of the exponentially decaying pion cloud. 
A second consequence is the absence of linear pion radiation, and one may
wonder whether there exists classical radiation at all in this model.
The answer is probably yes, although the study of radiation is inherently
nonlinear in compacton-supporting models of this type (the field equations
remain nonlinear in the weak-field limit). 
The simplest way to find some indications of radiation is the study of
rotating solitons. In the standard Skyrme model (with a nonzero pion mass), it
is found that rotating solutions exist for not too large angular velocities
but cease to exist if the angular velocity exceeds a certain limit. The reason
for this behaviour may be understood easily from the linearized weak-field
analysis. If the corresponding angular frequency is too large (essentially
larger than the pion mass), then the formal solution is oscillatory instead of
exponentially decaying, and so has infinite energy. Physically this is
interpreted as the onset of pion radiation at that frequency. 
So one may wonder what happens for rotating solitons in the BPS Skyrme model.
Unfortunately, the field equations in this case can no longer be reduced to an
ordinary differential equation. There exists, however, a baby
Skyrme version of the BPS Skyrme model in one dimension lower, where
the dimensional reduction of a rotating baby Skyrmion ansatz to an ODE is
possible and has been performed in \cite{GP}. The result is as follows: the
rotating baby Skyrmion solution exists and can be found exactly if the angular
velocity remains below a certain critical value. 
It remains compact, and its radius even shrinks with the angular velocity
(although the moment of inertia increases, as one would expect).
For frequencies above the
critical value, on the other hand, a solution does not exist. This may be
viewed as an indication that radiation will set in also for a sufficiently
fast rotating BPS Skyrmion, although we repeat that radiation for compactons
is an inherently nonlinear and, therefore, complicated problem. 
\\
Finally, let us repeat that in the large $N_c$ expansion the 
meson-meson couplings 
are of order $1/N_c$. Hence, mesons become free and non-interacting at 
$N_c=\infty$ \cite{thooft}. From this perspective, the BPS Skyrme model 
(at $N_c=\infty$) provides an acceptable, although rather radical result. 
Namely, mesons are not only non-interacting, they 
disappear completely from the particle spectrum, their only remnants being some
collective (solitonic) excitations, and the chiral symmetry breaking aspects 
of pion dynamics taken into account by
the potential. This fact is crucial to 
cure the unwanted strong forces at intermediate range in the Skyrme model 
at large $N_c$.

\vspace*{0.1cm}
{\it Summary.} 
As announced previously, 
we found that the classical model already describes rather
well some features of the liquid drop model of nuclei.
These classical results are
probably more trustworthy for not too small nuclei, because 
\begin{itemize}
\item[i)] the contribution of the pion cloud (which is absent in our model) 
to the size of the nucleus is of lesser significance for larger nuclei. We
remind that in addition to the core of a nucleus (with a size which grows
essentially with the third root of the baryon number) a surface term is known
to exist for physical nuclei
whose thickness is essentially independent of the baryon number.
\item[ii)] the description of a nucleus as a liquid drop of nuclear matter is
  more appropriate for larger baryon number.
\item[iii)] the contribution of (iso-) rotational quantum 
excitations to the total mass of
a nucleus is smaller for larger nuclei, essentially because of the larger
moments of inertia of larger nuclei.   
\end{itemize}
We will find further indication for this behaviour in the next subsection,
where a rigid rotor quantization of the (iso-) rotational degrees of freedom is
performed for the $B=1$ nucleon. Indeed, as we shall see,
both the corresponding
(iso-) rotational excitations and the (missing) pion cloud will be of some
importance in this case.
%%%%%%%%%%%%%
\subsection{Quantization}
%%%%%%%%%%%%%
Let us now discuss the issue of quantization of the BPS Skyrme model. As the 
model is rather unusual, not containing the quadratic, sigma 
model kinetic part, one might doubt whether the quantization procedure can 
be performed. However, the sextic derivative term used in the construction, 
the square of the pullback of the volume on the target space,  is a very 
special one. It is the unique term with sextic derivatives which leads to a 
Lagrangian of second order in time derivatives. Therefore, we deal with a 
Hamiltonian of second order in time derivatives and the system can be 
quantized in the standard manner.
\\
We want to perform the semiclassical quantization about a soliton solution in
the same way it is performed for the standard Skyrme model. Let us recall that
for the nonzero or vibrational modes, the semiclassical quantization consists
in a quantization  of the quadratic oscillations about the classical
solution. These oscillations presumably just amount to renormalizations of the
couplings of the theory and therefore may be taken into account implicitly by
fitting the model parameters to their physical values. The zero mode
fluctuations related to the symmetries, on the other hand, cannot be
approximated by quadratic fluctuations and have to be treated by the method of
collective coordinates. In principle, one collective coordinate has to be
introduced for each symmetry transformation of the model which does not leave
invariant the soliton about which the quantization is performed. 
Here, nevertheless, we only shall consider the collective coordinate
quantization of the rotational and isorotational degrees of freedom. 
The physical reason for this restriction is, of course, the fact that the
excitational spectra of nuclei are classified exactly by the corresponding
quantum numbers of spin and isospin. A more formal justification of this
restriction could be, for instance, that the additional collective coordinates
do not provide discrete spectra of excitations but, instead, just renormalize
the coupling constants, like the vibrational modes do. A definite answer to
this question would require a more detailed investigation of the full moduli
space of the theory, where all the infinitely many symmetries are taken into
account. This is probably a very difficult problem which is beyond the scope
of the present paper. A second justification consists in the assumption that,
in any case, the given model is just an approximation, whereas a more detailed
application to the properties of nuclei requires the inclusion of additional 
terms in the Lagrangian which, although being small in some sense, 
have the effect of breaking the symmetries down to the ones of the
standard Skyrme model.    
\\
We start from the classical, static field configuration 
$U_0$ found in Section 2.5. 
For simplicity, we only consider the hedgehog configuration with baryon number
$B=1$. This configuration is invariant under a combined rotation in base and
isotopic space, therefore, it is enough to introduce the collective
coordinates of one of the two. Allowed excitational states will always have
the corresponding quantum numbers of spin and isospin equal, as a consequence
of the symmetries of the hedgehog.
Following the standard treatment, we introduce the collective 
coordinates of the isospin by including a time-dependent iso-rotation of the 
classical soliton configuration
\begin{equation}
U(x)=A(t)U_0(x)A^{\dagger}(t),
\end{equation}
where  $A(t)=a_0+ia_i\tau_i \in SU(2)$ and $a_0^2+\vec{a}^2=1$. Inserting this
expression into the Lagrangian, we get
\begin{equation}
L=-E_0 +{\cal I} \, {\rm Tr} [ \partial_0 A^{\dagger} (t) \partial_0 A(t)]] 
\end{equation}
where the energy (mass) of the classical solution is
\begin{equation}
E_0=\frac{64\sqrt{2} \pi}{15} \mu \lambda
\end{equation}
and the moment of inertia is 
\begin{equation}
{\cal I}=\frac{4\pi}{3} \lambda^2 \int_0^{\infty} dr (\sin^4\xi \xi'^2_r)=
\frac{4\sqrt{2} \pi}{3} \sqrt{\lambda \mu} \left( \frac{3\lambda}{\sqrt{2} 
\mu } \right)^{2/3} \gamma, 
\end{equation}
where
$$ \gamma =  \int_0^{\infty} dz (z^{2/3} \sin^4\xi \xi'^2_z)=4\int_0^{4/3} 
z^{2/3}\left(1-\left( \frac{3z}{4}\right)^{2/3} \right)dz = \frac{32}{35} 
\left( \frac{4}{3}\right)^{2/3}.$$ 
Then, finally 
\begin{equation}
{\cal I}= \frac{2^8 \sqrt{2} \pi}{15\cdot 7}  \lambda \mu 
\left( \frac{\lambda}{\mu}\right)^{2/3}.
\end{equation}
Introducing the conjugate momenta $\pi_n$ to the coordinates $a_n$ on SU(2) 
$\simeq S^3$ we get 
the Hamiltonian 
$$ H=E_0-\frac{1}{8{\cal I}} \sum_{n=0}^3 \pi_n^2 =
E_0-\frac{\hbar^2}{8{\cal I}} \sum_{n=0}^3\frac{\partial^2}{\partial a_n^2}
$$ 
where the usual canonical quantization prescription $\pi_n \to -i\hbar
\partial / \partial a_n$ has been performed. Finally we get
$$
H = E_0
+ \frac{\hbar^2 I^2}{2{\cal I}}=E_0+\frac{\hbar^2 S^2}{2{\cal I}},
$$
where $I^2$ is the isospin operator (the spherical laplacian on $S^3$).
We introduced $\hbar$ explicitly because later on we want to use units where
$\hbar$ is different from one.
Further, $S^2$ is the spin operator, and we took into account the equality of
spin and isospin for the hedgehog. It is interesting to note that the isospin
operator automatically allows for wave functions on $S^3$ both for integer
isospin (homogeneous polynomials of even degree) and half-odd integer isospin
(homogeneous polynomials of odd degree). 
\\ 
The soliton with baryon number one is quantized as a fermion. Concretely, the
nucleon has spin and isospin 1/2, whereas the $\Delta$ resonance has spin and
isospin 3/2, so we find for their masses 
\begin{equation}
M_N=E_0+\frac{3\hbar^2}{8{\cal I}}, \;\; M_\Delta=E_0+
\frac{15\hbar^2}{8{\cal I}} \;\; 
\Rightarrow \;\; 
M_\Delta-M_N=\frac{3\hbar^2}{2{\cal I}},
\end{equation}
which is exactly like the nucleon and delta mass splitting formula of the 
standard Skyrme model. The difference comes only from particular expressions 
for $E_0$ and $I$. These expressions may now be fitted to the physical masses
of the nucleon ($M_N = 938.9$ MeV) and the $\Delta$ resonance ($M_\Delta =
1232$ MeV), which determines fitted values for the coupling
constants. Concretely we get
$$
\lambda \mu = 45.70 {\rm MeV} \, , \qquad 
\frac{\lambda}{\mu} = 0.2556 {\rm fm}^3 
$$
where we used $\hbar = 197.3$ MeV fm. These may now be used to ``predict''
further physical quantities like, e.g. the charge radii of the nucleons.
For this purpose, we need the linear (i.e., per unit radius)
isoscalar and isovector charge
densities. These expressions have already been determined for a generalized
Skyrme model including the sextic term in \cite{ding}, so we just use these
results in the appropriate limit.
\\
For the isoscalar (baryon) charge density per unit $r$ we find
\begin{equation}
\rho_0 = 4\pi r^2 B^0= - \frac{2}{\pi} \sin^2 \xi \xi'_r 
\end{equation}
and for the isovector charge density per unit $r$
\begin{equation}
\rho_{1}=\frac{1}{{\cal I}} \frac{4\pi}{3} \lambda^2  \sin^4\xi \xi'^2_r.
\end{equation}
Then the electric charge densities for proton and neutron, $\rho_{E, p(n)}
= \frac{1}{2} (\rho_0 \pm \rho_{1})$ read
\begin{equation}
\rho_{e, p(n)}=\frac{2\sqrt{2}}{\pi} \frac{\mu}{\lambda} r^2 
\sqrt{1-\left( \frac{\mu}{2\sqrt{2} \lambda}\right)^{2/3}r^2} \left( 
1\pm  \frac{4\sqrt{2}\pi^2 \lambda \mu r^2}{3 {\cal I} }\sqrt{1-\left( 
\frac{\mu}{2\sqrt{2} \lambda}\right)^{2/3}r^2} \right).
\end{equation}
The corresponding isoscalar and isovector mean square electric radii are
\begin{equation}
<r^2>_{e,0}= \int dr r^2 \rho_0 = \left( \frac{ \lambda}{\mu} \right)^{2/3},
\end{equation}
\begin{equation}
<r^2>_{e,1}= \int dr r^2 \rho_1 = 
\frac{10}{9} \left( \frac{ \lambda}{\mu} \right)^{2/3}.
\end{equation}
Further, the isoscalar magnetic radius is defined as the ratio
\begin{equation}
< r^2>_{m,0} = \frac{\int dr r^4 \rho_0}{\int dr r^2 \rho_0} = \frac{5}{4}
\left( \frac{ \lambda}{\mu} \right)^{2/3}.
\end{equation}
With the numerically determined values of the coupling constants we find the
values for the radii displayed in Table \ref{table2}.  
\\
\begin{table}
\begin{center}
\begin{tabular}{|c|c|c|c|}
\hline
radius & experiment & BPS Skyrme & massive Skyrme   \\
\hline 
compacton & - &  0.897 & -  \\
electric isoscalar $r_{e,0}$ & 0.72 &  0.635 &  0.68  \\
electric isovector $r_{e,1}$ & 0.88 &  0.669  &  1.04  \\
magnetic isoscalar $r_{m,0}$ & 0.81 &  0.710 &  0.95  \\
\hline
$r_{e,1}/r_{e,0}$ & 1.222 & 1.054 & 1.529  \\
$r_{m,0}/r_{e,0}$ & 1.125 & 1.118 & 1.397  \\
$r_{e,1}/ r_{m,0}$ & 1.086 & 0.943 & 1.095  \\
\hline
\end{tabular}  
\caption{Compacton radius and some charge radii and their ratios
for the nucleon. The numbers
  for the massive Skyrme model  are from Ref. \cite{nappi2}.
All radii are in fm.} \label{table2}
\end{center}
\end{table}
In relation to Table \ref{table2}, some comments are appropriate. Firstly,
observe that in the BPS Skyrme model all radii are
bound by the compacton radius $R_0= \sqrt{2}(\lambda /\mu)^{(1/3)}$. 
This bound holds because all radii can be expressed as 
moments of densities $(\int dr r^n \rho_i)^{(1/n)}$ where $\rho_i $ is a
density normalized to one. Secondly, all radii in the BPS Skyrme model
are significantly
smaller than their physical values, as well as significantly smaller than the
values predicted in the standard massive Skyrme model. This, however, has
to be expected, because we know already that the pion cloud is absent in the
BPS Skyrme model, and the densities strictly go to zero at the
compacton radius. We also display the ratios of some radii for the following
reason. If the deviations of the BPS Skyrme model radii 
from their physical values are mainly due to the same ``systematic error'' 
(the absence of the pion cloud in the model), then we expect that this 
``systematic error'' should partly cancel in the ratios. 
This is precisely what happens. The
errors in the radii themselves are of the order of 30\%, whereas the errors in
the ratios never exceed 15\%, providing us with a nice consistency check for
our interpretation of the model.  

Finally, let us display the numerical results for the magnetic moments of the
proton and neutron. The corresponding expressions are
\begin{equation}
\mu_{p(n)} = 2M_N \left( \frac{1}{12 {\cal I}} <r^2>_{e,0} + (-) \frac{{\cal
    I}}{6\hbar^2} \right) ,
\end{equation}
and the resulting numerical values are given in Table 3.
\\
\begin{table}
\begin{center}
\begin{tabular}{|c|c|c|c|}
\hline
 & experiment & BPS Skyrme & massive Skyrme  \\
\hline 
$\mu_p $ & 2.79 & 1.918 & 1.97   \\
$\mu_n $ & -1.91 &  -1.285 &  -1.24  \\
$|\mu_p / \mu_n |$ & 1.46 &  1.493  &  1.59  \\
\hline
\end{tabular}  
\caption{Proton ande neutron magnetic moments. The numbers
  for the massive Skyrme model are from Ref. \cite{nappi2}.} \label{table3}
\end{center}
\end{table}
The quality of the values is comparable to the case of the standard massive
Skyrme model, so the absence of the pion cloud apparently does not have such a
strong effect on the magnetic moments.

%%%%%%%%%%%%%%%%%%%%%%%%%%%%%%%%%%%%%%%%%%
\section{Conclusions}
%%%%%%%%%%%%%%%%%%%%%%%%%%%%%%%%%%%%%%%%%%
In this work we proposed an integrable limit within the space of 
generalized Skyrme models which 
we called {\it the BPS Skyrme model}. The model consists of two terms: the 
square of the pullback of the target space volume form (or, equivalently, of
the topological density) and 
a non-derivative part, i.e., a potential. Both terms are required to guarantee 
the stability of static solutions. This theory possesses rather striking 
properties. It is integrable, that is, there are infinitely many conserved 
charges. It is also solvable for any form of the potential with solutions 
given by quadratures. 
Further, all
solutions are of the BPS type. They obey a first order differential 
equation and saturate a Bogomolny bound. 
These properties provide the model with some independent mathematical interest
of its own, although in this paper our main concern dealt with its possible
relevance as a low-energy effective field theory for strong interaction
physics and for the phenomenology of nuclei. 
\\
Firstly, let us emphasize again the 
possible relevance of the BPS Skyrme model in
the limit of a large number of colors $N_c$ of the underlying QCD type
theory. Indeed, as was pointed out, e.g., in \cite{larry}, some problems
of the standard Skyrme model when applied to QCD like theories become more
severe in the large $N_c$ limit. For instance, in the Skyrme model rather
strong forces of order $N_c$ are generated between nuclei, and the ground
state of sufficiently high baryon number tends to be a Skyrmion crystal with
binding energies again of order $N_c$. Both of these findings are in conflict
with lattice simulations and with known properties of physical nuclei,
respectively. On the other hand, both of these issues are absent in the BPS
Skyrme model (there are no long range forces and no binding energies). So one
might speculate that the BPS Skyrme model provides more accurate results as an
effective field theory in the large $N_c$ limit. Unfortunately, however, there
is no obvious large $N_c$ limit which would
produce just the BPS Skyrme model as
its leading order, so the rather good large $N_c$ properties of the model must
be due to some more subtle mechanism. A better theoretical understanding of
the conditions under which the BPS Skyrme theory provides a reasonable limit
as an effective field theory for large $N_c$ QCD like theories would be
highly desireable. It might, for instance, happen that the two terms are
enhanced by two different physical mechanisms, where the sextic term is
related to some collective or topological excitations, whereas the potential is
related, e.g., to the chiral quark condensate of QCD.
\\
Secondly, in our attempts for direct physical applications of the BPS
Skyrme theory,
analogously to what is usually done for the standard Skyrme theory, 
we found that already the classical
solitons in the BPS Skyrme model have properties 
which make this idea worthy of discussion. The mass (energy) of the solitons 
is proportional to the baryon charge $E \sim |B|$, which as we know from 
experimental data, is, to a good degree, a feature of physical nuclei. 
Moreover, also the radii of the solitons follow the standard experimental 
law $R_0 \sim |B|^{1/3}$. Additionally, the energy density is of nucleus 
type. The linear energy-charge relation is valid for all potentials. Thus, 
it is, in principle,
possible to specify a particular potential by fitting the resulting 
energy density to the experimental data. We find it quite intriguing that 
this simple BPS Skyrme model gives significantly better approximations - at
least on a purely classical level - to 
the experimental energies (and densities) of baryons than the original 
Skyrme model (and its generalizations).  However, as we neglected the 
standard kinetic term for the chiral field, there are no obvious 
pseudo-scalar degrees of freedom ($\eta, \vec{\pi}$). Fortunately, the 
model contains  terms maximally second order in time derivative, and 
therefore, it has the standard time dynamics and hamiltonian interpretation. 
Thus, it is possible to investigate interactions of such solitons and 
identify effective forces between them (and effective degrees of freedom 
carrying the interactions).    
\\
Thirdly, for an application to the physics of nuclei 
the issue of quantization of the BPS Skyrme 
solitons certainly has to be investigated. 
It results that the standard semi-classical rigid rotor quantization of the 
rotational and iso-rotational degrees of 
freedom of a Skyrmion can be performed in a completely equivalent fashion for
the BPS Skyrme model. We explicitly did this rigid rotor quantization for the
$B=1$ Skyrmion (nucleon), and the numerical results conform well with the
physical interpretation of the BPS Skyrme model when applied to nuclei. For
instance, the absence of the pion cloud is clearly reflected in the too small
values for the resulting charge radii, whereas ratios of these radii, where
the pion cloud effect partially cancels, agree quite well with their
experimental counterparts.  We did not explicitly calculate the (iso-)
rotational excitation spectra of Skyrmions with higher baryon charge, but
as we still deal with exact solutions having continuous 
symmetries, one may expect that this task should not be too difficult. 
\\
On the
other hand, for the standard Skyrme theory and its generalizations, higher
Skyrmions have rather complicated discrete symmetries and are known only in 
numerical form, so their quantization is a 
rather complicated procedure. Nevertheless, recently 
the rotational and
isorotational excitations of the rigid rotor quantization of 
the solitons of the
standard massive Skyrme model have been applied quite successfully to the
corresponding spectra of excitations of light nuclei \cite{sutcliffe1}. 
As the solutions in the
standard Skyrme model are sometimes quite different from ours, one might think
that this fact casts some serious doubts on the applicability of our model to
the phenomenology of nuclei. Here we just want to point out that this does not
have to be the case. In fact, the information which is most important for the
spectra of excitations consists in the {\em symmetries} of the solitons, and
not in the full dynamical contents of the soliton solutions. These symmetries
determine the Finkelstein--Rubinstein constraints on the allowed excited
states and, therefore, the spectra of excitations for each baryon number.
Further, the solutions in our model typically have higher symmetry due to the
special properties of this model. 
\\
As a consequence, the following picture is
quite plausible. Our model as it stands already describes quite well some bulk
properties of nuclei like masses and charge and energy densities. A more
detailed description does require the addition of further terms, but these
will be small in some sense (e.g. their contribution to the total mass is
small). On the other hand, these additional terms will break the symmetries of
the resulting soliton solutions, and these solutions probably have the
symmetries of the standard Skyrme model, and, consequently, their spectra of
excitations. 
If this symmetry breaking is small, then the spectral lines should still show
some approximate degeneracy, that is, some spectral lines should be spaced
more narrowly than others. A detailed investigation of this issue 
is beyond the scope of the present paper and will be presented in future
publications. 
Of course, in the simplest baryon number one case (the hedgehog), 
the symmetries and the excitational spectra coincide. 
\\
There are certainly some further applications of the BPS Skyrme model beyond
the realm of nuclear physics. One may, for instance,
consider it as a laboratory for skyrmions of the standard Skyrme model.
The BPS model 
allows for the analytical investigation of problems which can be studied
only by advanced 3D numerical simulations in the original Skyrme theory.
Or one may consider the BPS model as a lowest order approximation for more
complicated generalized Skyrme models and calculate the properties of the
latter by a kind of perturbative expansion about the BPS model.
These issues are, however, beyond the scope of the present work and shall be
investigated elsewhere.
\\
Summarizing, we believe that we have identified and solved
an important submodel in the
space of Skyrme-type effective field theories, 
which is singled out both by its
capacity to reproduce qualitative properties of the liquid drop
approximation of nuclei and by its unique 
mathematical structure.
The model directly relates the nuclear mass to the topological charge, and it
naturally provides both a finite size for the nuclei and the liquid drop
behaviour, which probably is not easy to get from an effective field
theory. (One wonders whether it is also possible to get the surface
contribution to the energy of the nuclear drop model from an effective field
theory, as the BPS Skyrme model does for the volume contribution.) 
So our model solves a conceptual problem by expliclity deriving said
properties from a (simple and solvable) effective field theory.
Last not least, our exact solutions might provide a calibration for the 
demanding numerical computations in physical applications of more 
general Skyrme models.

{\em Note added:} After finishing this article we became aware of a simultaneous paper \cite{Bon-Mar},
where the authors use a version of the BPS Skyrme model with a different potential to describe the binding energies of higher nuclei. Concretely, they first calculate the exact static soliton solutions plus the (iso-) rotational energies in the rigid rotor quantization for general baryon charge $B=n$ for the spherically symmetric ansatz (Section 2.4 in our paper). Then they allow for small contributions to the total energies from the quadratic and quartic Skyrme terms and fit the resulting binding energies to the experimental binding energies of the most abundant isotopes of higher nuclei, assuming, as is usually done, that these correspond to the states with the lowest possible value of the isospin. The resulting  agreement between calculated and experimentally determined masses and binding energies is impressive, lending further support to the viability of the BPS Skyrme model as the leading contribution to an effective theory for the properties of nuclear matter. 

\section*{Acknowledgements}

C.A. and J.S.-G. thank the Ministry of Science and Investigation, Spain 
(grant FPA2008-01177), and
the Xunta de Galicia (grant INCITE09.296.035PR and
Conselleria de Educacion) for financial support.
A.W. acknowledges support from the
Ministry of Science and Higher Education of Poland grant N N202
126735 (2008-2010).  Further, A.W. thanks M.A. Nowak and L. McLerran, and
J.S.-G. thanks M. Asorey, J.L. Cortes, J.V.G. Esteve and
V. Vento for interesting discussions.

 \end{document}